\documentclass[10pt,twocolumn,letterpaper]{article}

\usepackage{cvpr}              

\usepackage[dvipsnames]{xcolor}

\definecolor{cvprblue}{rgb}{0.21,0.49,0.74}
\usepackage[pagebackref,breaklinks,colorlinks,citecolor=cvprblue]{hyperref}

\def\paperID{*****} 
\def\confName{CVPR}
\def\confYear{2024}

\title{High-Resolution Image Translation Model Based on Grayscale Redefinition}

\author{Xixian Wu\\
Nanjing University of Science and Technology\\
NJUST, Nanjing, Jiangsu Province, China\\
{\tt\small xixianwu@njust.edu.cn}
\and
Dian Chao\\
Nanjing University of Science and Technology\\
NJUST, Nanjing, Jiangsu Province, China\\
{\tt\small chaodian@njust.edu.cn}
\and
Yang Yang*\\
Nanjing University of Science and Technology\\
NJUST, Nanjing, Jiangsu Province, China\\
{\tt\small yyang@njust.edu.cn}
}

\begin{document}
\maketitle
\begin{abstract}
Image-to-image translation is a technique that focuses on transferring images from one domain to another while maintaining the essential content representations. In recent years, image-to-image translation has gained significant attention and achieved remarkable advancements due to its diverse applications in computer vision and image processing tasks. In this work, we propose an innovative method for image translation between different domains. For high-resolution image translation tasks, we use a grayscale adjustment method to achieve pixel-level translation. For other tasks, we utilize the Pix2PixHD model with a coarse-to-fine generator, multi-scale discriminator, and improved loss to enhance the image translation performance. On the other hand, to tackle the issue of sparse training data, we adopt model weight initialization from other task to optimize the performance of the current task.
\end{abstract}    
\section{Introduction}
\label{sec:intro}

As deep neural networks continue to evolve, significant strides have been achieved in numerous fields, including object detection, multimodal classification~\cite{yang2019semi,yang2018complex,yang2019deep}, cross-modal retrieval~\cite{yang2021rethinking,yang2024alignment}, and so on. In addition, some work is dedicated to implementing deep neural networks in specific applications, such as recommendation systems~\cite{yang2023contextualized}, rumor detection~\cite{yang2023deep}, and so forth. Given the significant role visual information plays in human perception, a challenge emerges in effectively transferring information across distinct visual domains, a feat known as image translation.
Image translation enables the preservation of content and the modification of style by transforming images from the source domain to the target domain. It is widely applicable in various sectors such as facial attribute editing, scene style transformation, and many more.
Despite a series of advancements in the field of image translation over the past few years, the complexity of the task still presents many unresolved challenges. 

Based on the framework of generative adversarial networks, such as pix2pix~\cite{isola2017image}, cycleGAN~\cite{zhu2017unpaired}, and pix2pixHD~\cite{wang2018high}, remarkable capabilities have been demonstrated in the field of image translation. These models adopt a generator and discriminator structure , where the generator uses an encoder decoder structure which has achieved success in many fields~\cite{yang2022exploiting} to translate the source domain image into the target domain image, and the discriminator distinguishes whether the image is real or fake. In the game between the two, the generator gradually generates more realistic target domain images.

However, for high-resolution image pairs, GANs often produce artifacts and color distortions in the generated images. To solve this problem, different strategies are used for different tasks. For the RGB2IR task, we propose an innovative technique aimed at enhancing the quality of the generated IR images. When tackling SAR2EO, SAR2RGB, and SAR2IR tasks, we utilized the Pix2PixHD model. Due to the size limitations of SAR2RGB and SAR2IR datasets, we utilized the pre-trained weights of the Pix2PixHD model on the SAR2EO dataset for further training on SAR2RGB and SAR2IR images. In the end, our approach performed outstandingly in all four tasks, achieving a final score of 0.32, and ranked first on the leaderboard of CVPR PBVS 2024.


\section{Related Work}
\label{sec:Related Work}

\subsection{Generative adversarial networks}

Generative Adversarial Networks (GANs)~\cite{goodfellow2014generative} is a type of unsupervised deep learning framework, first proposed by Ian Goodfellow in 2014. This framework primarily consists of two neural networks: a generator and a discriminator, which are trained via a game-theoretic setup.

The generator is responsible for generating new data samples. Its goal is to "deceive" the discriminator such that it cannot differentiate between the samples produced by the generator and those from the real world. The generator takes as input random noise, which then undergoes a series of nonlinear transformations (usually made up of convolutional layers and up-sampling layers) to generate fake samples identical in shape to the real data.

The discriminator, on the other hand, is designed to be a binary classifier. Its task is, given a sample, to determine whether it is generated by the generator (i.e., fake) or taken from the training set (i.e., real). The ideal situation is when the discriminator cannot differentiate the authenticity, given that the fake samples and real samples share identical probability distributions.

The training process of GANs can be interpreted as a kind of "zero-sum game", in which the generator and discriminator compete against each other to reach a dynamic equilibrium point. Researchers usually strike this adversarial balance by adjusting the loss function and optimization algorithm.

GANs have achieved significant success in many fields, including but not limited to image synthesis, super-resolution, style transfer, sample generation, and text generation.

However, the training of generative adversarial networks still poses significant challenges. Mode collapse~\cite{metz2016unrolled} and training instability~\cite{arjovsky2017towards} are the hot issues of current research. 
Some works aims to find more efficient and stable ways for adversarial training to address these issues~\cite{liu2019spectral}~\cite{srivastava2017veegan}.

\subsection{Image-to-image translation}

Within the realm of computer vision, image-to-image translation has emerged as a dynamic and crucial field. The primary objective of this discipline is to establish the mapping relationship between source domain images and target domain images. This swiftly developing area of study aims to translate the learned mapping to new images that hold a similar relationship such that any image of a particular style, condition, or genre can be transformed into another.

However, such progress has not been without considerable challenges. One of the key issues revolves around the data used for training. Paired image translation, where the training dataset involves photographic pairs of the same subject from different perspectives or conditions, is generally easier due to the definite correspondence between the images. However, obtaining image pairs with precise pixel-to-pixel mapping is highly impractical, and often the datasets exhibit no one-to-one correspondence, making the training more difficult.

Moreover, the successful translation of high-resolution images via adversarial methods continues to pose significant challenges. Training instability and optimization issues remain persistent hurdles – as Chen and Koltun~\cite{chen2017photographic} pointed out. Oftentimes, even if high-resolution images can be synthesized, they may lack fine details and accurate textures.

Various methodologies have been proposed to tackle these issues. Lately, adversarial loss-based techniques for image translation have gained significant traction. The pix2pix framework, for example, uses image-conditional GANs~\cite{mirza2014conditional} for multi-faceted applications like transforming Google maps into satellite views or generating cat images from user sketches.

Unpaired image translation, mapping images between two or more domains where instances take no precise correspondences, has also been explored. The introduction of multiscale generators and discriminators, as well as novel objective functions, remains influential, aiding the stabilization of conditional GANs during training. This innovative breakthrough also serves to generate high-resolution images more effectively.

Despite these advancements, thorough research and development are still imperative in ensuring successful image translation, making this a fascinating, albeit challenging, field of study.

\subsection{Conditional GANs}

Conditional Generative Adversarial Networks (Conditional GANs)~\cite{mirza2014conditional} are a variant of GANs that incorporate auxiliary information to guide the generation process. This can potentially enhance the usability and controllability of generated images.

In Conditional GANs, the generator and discriminator are conditioned on certain auxiliary information. This brings an added level of complexity to the generation process as results now correspond to specified conditions.

One of the primary applications of Conditional GANs is in image-to-image translation tasks where they leverage input-output image pairs to guide the translation process.

\subsection{pix2pix}
Pix2pix~\cite{isola2017image} presents a compelling exploration into conditional adversarial networks as a comprehensive solution to the various challenges of image-to-image translation. This presents a contrast to the conventional GANs, which are primarily trained to map from a random noise distribution to a distribution of real images. Pix2pix, however, is architected to evolve from input images to output images. Essentially, this means that Pix2pix is capable of transforming one feasible representation of a scene into another representation.

Historically, each of these image translation tasks would require separate, dedicated machinery. But Pix2pix pushes the envelope by proposing the use of image-conditioned generation adversarial networks as a one-size-fits-all solution for image-to-image translation. The Pix2pix approach, thus, stands out due to its potential universality in solving this diverse set of problems.




\section{Method}
\subsection{Overall Framework}

\begin{figure*}[t]
  \centering
  \includegraphics[width=\textwidth]{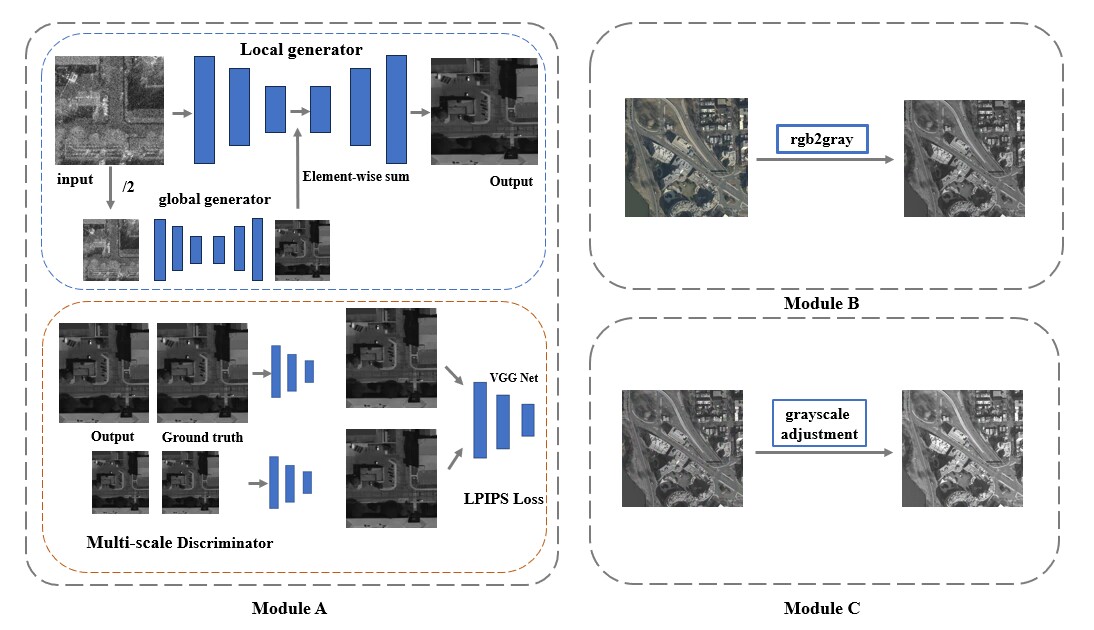}
  \caption{The main structure of Our model.}
  \label{fig:your-label}
\end{figure*}

This solution primarily consists of three modules. Firstly, we utilize the Pix2pixHD model based on a pre-trained network with modifications to certain losses. Following that, there is a module for converting RGB images to grayscale, and finally, a module for adjusting grayscale images. We combine these modules in different ways for various tasks, achieving either first or second place in performance across all tasks (except for the SAR2RGB task, where we consistently achieve first place on the test set for other tasks).

\subsection{Modifying the loss of Pix2pixHD}
Firstly, according to the competition requirements, the evaluation metric is:
\begin{equation}
\text{Final Score} = \frac{\frac{2}{\pi}\arctan(\text{FID}) + \text{LPIPS} + \text{L2}}{3}
\label{eq:final_score}
\end{equation}

To enhance the generated images' scores in the final evaluation metric, we incorporate this metric as a loss during the training process of Pix2pixHD. Specifically, the calculation method for LPIPS is as follows:
\begin{equation}
F_t = \text{VGG}(I_t), \quad F_g = \text{VGG}(I_g)
\end{equation}
\begin{equation}
\text{Diff} = \|F_t - F_g\|_2
\end{equation}
\begin{equation}
\text{LPIPS} = \text{Aggregation}(\text{Diff})
\end{equation}

Initially, features are extracted from both the generated and real images using the VGG network, denoted as $\text{F}_g$ and $\text{F}_t$, representing features extracted from the generated and real images, respectively. Subsequently, the difference between the two feature vectors is computed using the Euclidean distance. Finally, the ultimate LPIPS score is obtained by evaluating the difference scores.
\begin{equation}
\text{L2}(\text{I}_g, \text{I}_t) = \sqrt{\sum_{i,j} ({I_g}_{i,j} - {I_t}_{i,j})^2}
\end{equation}

The calculation of the L2 norm measures the difference between corresponding pixels in the generated and real images. The greater the difference between the two images, the larger the value of the L2 norm.
\begin{equation}
\text{Loss}_{all} = \text{Loss}_{ori} + \text{LPIPS} + \text{L2}
\end{equation}

In the end, the total loss of the modified Pix2pixHD model is the sum of the original Pix2pixHD model loss, LPIPS loss, and L2 loss.

\subsection{Using pre-trained weights of SAR2EO}
Due to the constrained size of the SAR2RGB and SAR2IR datasets, we opted to utilize the Pix2PixHD model pretrained on the SAR2EO dataset. The pretrained model was directly instantiated and subjected to further training on SAR2RGB and SAR2IR images. This strategic approach was undertaken to augment the model's proficiency in generating finer details, thereby leading to an enhancement in overall performance pertaining to these specific tasks.
\subsection{Convert to grayscale}
During the model training process, we observed that in the RGB2IR task trained by the Pix2pixHD model, the generated IR images, while possessing high resolution, exhibited significant discrepancies in terms of grayscale when compared to real IR images. Consequently, we employed a straightforward yet effective approach to bring the generated IR images closer to the realism of the authentic counterparts.

Specifically, we computed the RGB values of each pixel in the image relative to black, resulting in an array representing color differences. Subsequently, we calculated the average of color differences across the three channels, yielding a grayscale image array. Each pixel value in this array indicates the average difference between the color at the corresponding position and black. Finally, we converted the grayscale image array back into an image, thus generating high-quality IR images.
\begin{figure}
  \centering
  \includegraphics[width=0.9\linewidth]{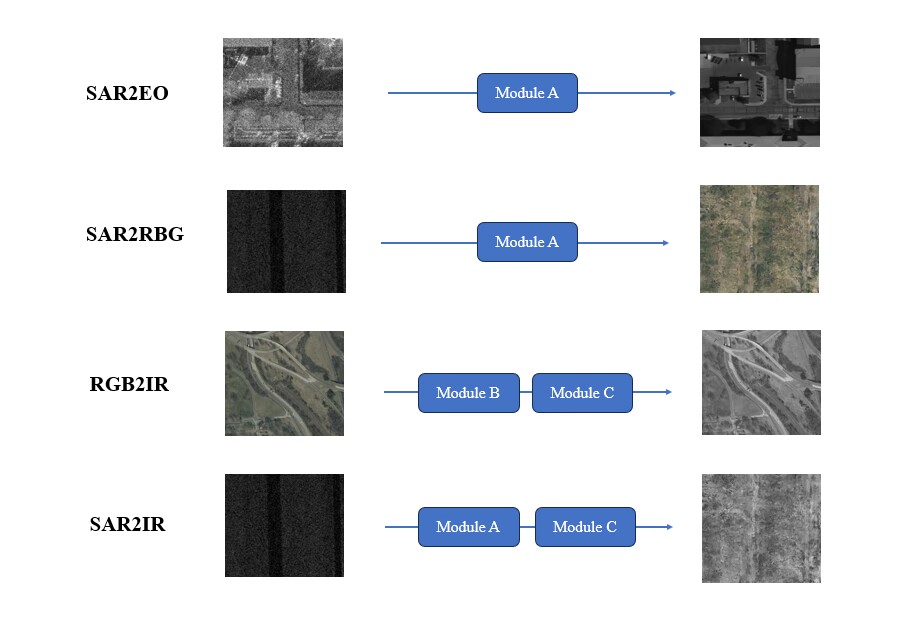}
  \caption{The different combination of modules for different tasks}
  \label{fig:your-label}
\end{figure}
\subsection{Gray density reconstruction}
To further refine the grayscale nuances of the generated images, we adjust the grayscale of the image by multiplying each element in the grayscale image array by an intensity factor. Subsequently, the obtained result is constrained within the range of 0 to 255, truncating values beyond this range. The adjusted image is then converted back into an image format. This step ensures that the pixel values of the adjusted image remain within a reasonable range, avoiding undesirable artifacts resulting from brightness adjustments.

Through iterative adjustments based on the generated results from the training set, we ultimately determined that, for the RGB2IR task, setting the intensity factor to 1.3 yielded optimal results. In the SAR2IR task, the intensity factor was set to 1.15 for optimal performance.
\subsection{Module combination for different tasks}
We have established distinct module combinations for different tasks. For the SAR2EO task, only the Pix2pixHD module with modified loss functions is employed.

The SAR2RGB task utilizes a Pix2pixHD model pretrained with SAR2EO task weights as the initial set of weights.

In the RGB2IR task, the initial step involves transforming RGB images into grayscale, followed by adjustments to the grayscale.

For the SAR2IR task, the SAR2RGB model is employed to convert RGB images, followed by grayscale adjustments to transform them into IR images.

\begin{table*}[t] 
  \centering
  \begin{tabular}{@{}cccccccc@{}}
    \toprule
    User & Combined Score & SAR2EO & SAR2RGB & RGB2IR & SAR2IR \\
    \midrule
    \textbf{NJUST-KMG} & \textbf{0.32} & \textbf{0.08} & 0.55 & \textbf{0.16} & \textbf{0.51}  \\
    USTC-IAT-United & 0.33 & 0.10 & \textbf{0.54} & 0.17 & 0.52 \\
    up6 & 0.35 & 0.12 & 0.56 & 0.19 & 0.54 \\
    wangzhiyu918 & 0.36 & 0.11 & 0.54 & 0.22 & 0.55 \\
    hsansui & 0.40 & 0.10 & 0.57 & 0.36 & 0.58 \\
    lemonGJacky & 0.40 & 0.33 & 0.57 & 0.19 & 0.53 \\
    \bottomrule
  \end{tabular}
  \caption{PBVS 2024 Multi-modal Aerial View Imagery Challenges - Translation test set leaderboard.}
  \label{tab:example}
\end{table*}

\begin{figure*}[t]
  \centering
  \includegraphics[width=0.7\textwidth]{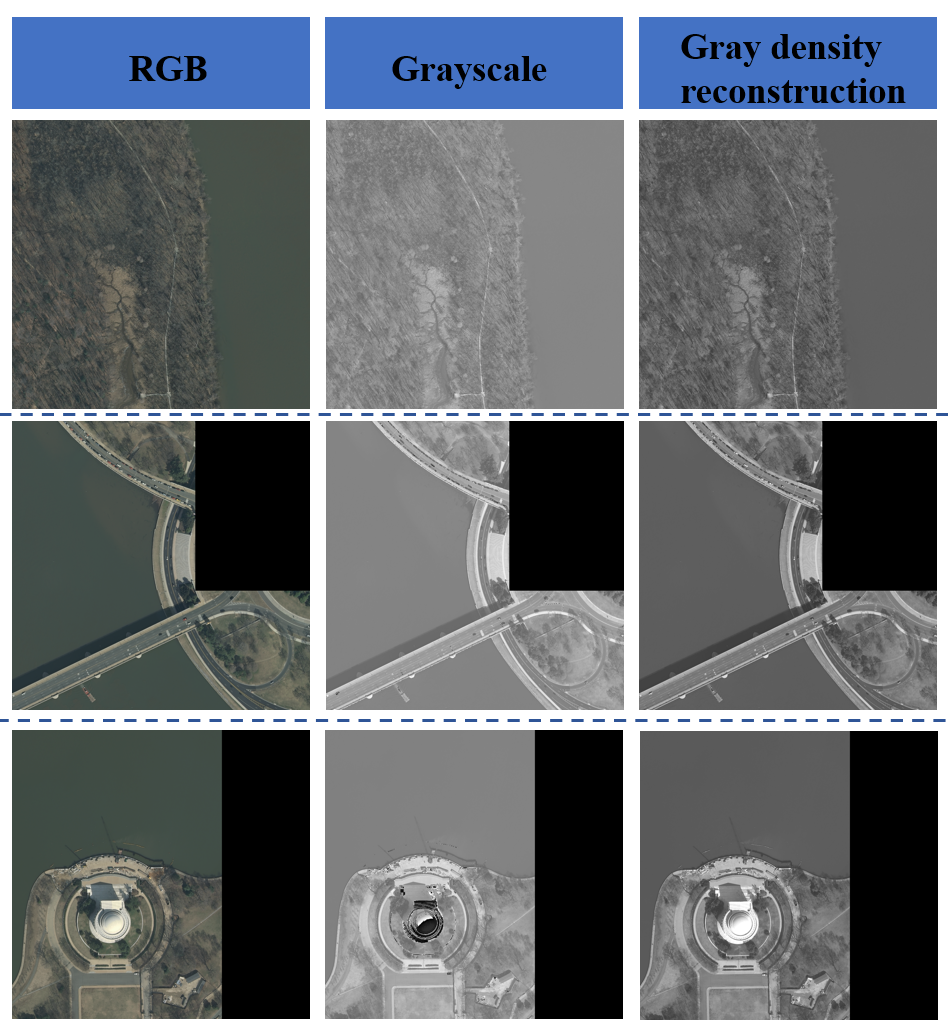}
  \caption{The process of transforming images from RGB to IR in the test set is illustrated in the RGB2IR conversion diagram.}
  \label{fig:your-label}
\end{figure*}

\section{Experiment}
In this section, we provide a brief overview of the dataset and evaluation metrics. Next, we present the implementation details of our approach. The performance of our approach on the dataset is evaluated quantitatively. Additionally, we conduct several ablation experiments to showcase the effectiveness of each component.
\subsection{Dataset}
The training data consists of Five different datasets, namely Train Data and Validation,Train Data UC Davis,Train Data Manhattan,Train Data Bingham,and Train Data Centerfield.

The training dataset includes RGB and IR images captured simultaneously and aligned in time. The SAR images at these locations were taken at different times from the RGB and IR images and maybe not aligned with RGB or IR images in space. Therefore, some objects may appear in RGB and IR images, but not in SAR images.
For each position, the provided training data includes at least one SAR image, at least one IR image, and at least one RGB image. Meanwhile, there may be different numbers of each image type available for training at different locations, and any additional images may be captured at different times.

In order to fully utilize the given dataset and prevent overfitting of images at specific locations, we adopted a random sampling strategy. For the images captured at each position, we randomly sample RGB, IR, and SAR images to form image pairs for training in SAR2IR, SAR2RGB, and RGB2IR tasks. For the SAR2EO task, use full Train Data (EO+SAR) data for training
\subsection{Implementation Details}
We conducted all experiments using two NVIDIA A6000 GPUs. Due to limited GPU memory, we adjusted the image resolution to 512*512 and 256*256 in some tasks. Specifically, for the SAR2EO task, we initially trained the model with a set learning rate for 30 epochs. Subsequently, we reduced the learning rate by half and continued training for an additional 10 epochs. As for the SAR2RGB task, we utilized Pix2pixHD weights from the pre-trained SAR2EO dataset, loading them and continuing the training process. This approach allowed the model to capture preliminary detailed features of the dataset's images.

Subsequently, the SAR2IR task was divided into two steps. First, we employed the trained SAR2RGB model to convert SAR images into RGB images. Then, by adjusting the grayscale, we transformed the RGB images into IR images.The RGB2IR task does not involve the use of the Pix2pixHD model. Instead, it relies solely on a two-stage grayscale conversion method designed by our team to transform RGB images into IR images.
\subsection{Evaluation metrics}
\textbf{L2 Norm:} recognized as Euclidean distance or L2 distance, stands as a widely used metric to evaluate the disparity between two vectors. It quantifies the vector's length, signifying the distance from the origin to the point it represents. In the field of image processing, the L2 Norm is a prevalent tool for calculating the pixel-wise divergence between two images.

\textbf{LPIPS:} constitutes a perceptual image quality metric designed to assess the similarity between two images, relying on the response of deep neural networks. Introduced in a research paper, LPIPS has demonstrated a strong correlation with human perception of image quality. The metric involves passing two images through a pre-trained deep neural network and calculating the distance between their feature representations, typically using the L2 Norm. The ultimate LPIPS score is derived by averaging these distances across multiple image patches.

\textbf{FVD:} is a metric designed to quantify the similarity between two sets of images. It relies on the distance between the feature representations of images, computed by a pre-trained deep neural network. The FVD metric involves calculating the mean and covariance of feature representations for both real and generated images. The Frechet distance, a measure of similarity between two multivariate Gaussian distributions, is then applied to compute the distance between the mean and covariance. A lower FVD score indicates a higher similarity between the two sets of images.
\subsection{Main results}
The results of our approach, as presented in Table 1, showcase a favorable performance when compared to other participants across various metrics. Also,the images generated by our model for different tasks are showcased in Fig 4. Notably, our method exhibits superior results in all evaluated aspects, except for the SAR2RGB task. It is worth noting that the potential reason for our model not achieving the state-of-the-art (SOTA) in the SAR2RGB task could be attributed to the decision to resize the images during training, resulting in a lower resolution. This resizing might have limited the model's exposure to essential details, potentially hindering its ability to achieve optimal training. Nonetheless, the overall performance of our method across diverse tasks underscores the effectiveness of the approaches adopted in this study.
\subsection{Ablation studies}
Table 2 presents the performance of different models in our ablation experiments. It is noteworthy that the "Combined Score" represents the final score after integrating various methods, while individual metrics such as SAR2EO, SAR2RGB, RGB2IR, and SAR2IR provide scores for specific tasks. The Pix2PixHD-c model, which represents the modified Pix2pixHD model, achieves a comprehensive score of 0.44. The Grayscale model is not applicable for certain tasks but excels with a score of 0.16 in the RGB2IR task and 0.53 in the SAR2IR task. The Pretrained model demonstrates outstanding performance in the SAR2RGB task with a score of 0.55. The "All Combined" approach, considering all models, obtains a comprehensive score of 0.32. These findings offer a comprehensive understanding of how different models contribute to the overall performance across diverse tasks.
\begin{table*}[t] 
  \centering
  \begin{tabular}{@{}cccccccc@{}}
    \toprule
    Model-name & Combined Score & SAR2EO & SAR2RGB & RGB2IR & SAR2IR \\
    \midrule
    Pix2PixHD-c &0.44 & 0.08 & 0.56 & 0.55 & 0.58\\
    Grayscale & / & / &  / & 0.16 & 0.53\\
    Pretrained & / & / & 0.55 & / & 0.55\\
    All Combined & 0.32 & 0.08 & 0.55 & 0.16 & 0.51\\
    \bottomrule
  \end{tabular}
  \caption{The impact of different models on the results is assessed, with these metrics evaluated on the test set. Pix2PixHD-c denotes the scores on the test set for the Pix2PixHD model with modified loss. Grayscale represents the scores after two-stage grayscale adjustments. Pretrained indicates the use of weights from the Pix2PixHD model trained on the SAR2EO dataset. All Combined signifies the scores obtained by combining these methods.}
  \label{tab:example}
\end{table*}

\begin{figure*}[t]
  \centering
  \includegraphics[width=0.9\textwidth]{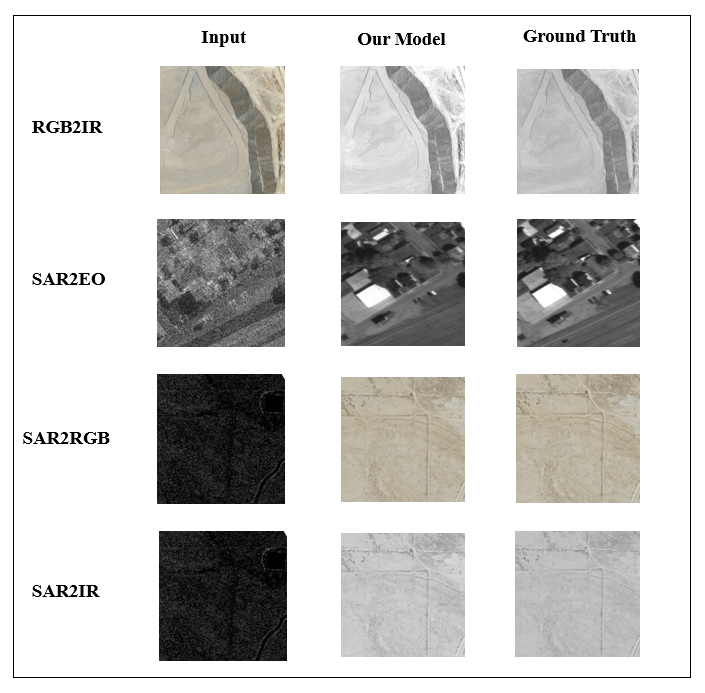}
  \caption{The comparison between results of our model and the ground truth.}
  \label{fig:your-label}
\end{figure*}

\section{Conclusion}

In our research, we present a pioneering approach to the transformation of RGB and SAR images into infrared (IR) counterparts, aiming to enhance the quality of resulting IR images. The initial step involves the conversion of RGB images into grayscale, followed by meticulous adjustments to pixel intensities using a dedicated intensity factor. This method stands out for its notable impact on significantly improving the overall quality of the generated IR images.

In addressing the SAR2EO, SAR2RGB, and SAR2IR tasks, we strategically employed the Pix2PixHD model for both training and testing phases. To contend with the substantial volume of data in the SAR2EO dataset, we implemented a nuanced adjustment of the learning rate during the final epochs of the training process, ensuring comprehensive and thorough dataset training. Meanwhile, due to the constrained sizes of the SAR2RGB and SAR2IR datasets, we capitalized on the Pix2PixHD model's pretraining on the SAR2EO dataset. This pretrained model was seamlessly integrated, undergoing further training specifically on SAR2RGB and SAR2IR images. This approach not only bolstered the model's capability to generate finer details but also yielded discernible improvements in performance across these targeted tasks.

To elevate the Pix2PixHD model's performance on the competition dataset, we introduced the L2norm evaluation metric and the LPIPS evaluation metric as integral components in the model's training process. This strategic integration was pivotal in enabling the model to attain higher levels of performance during the final scoring, underscoring its robustness and efficacy. It is worth noting that our approach consciously did not incorporate denoising as a specific enhancement strategy, differentiating our methodology from other contemporaneous strategies in the field. This deliberate choice underscores the uniqueness and effectiveness of our multi-faceted approach to image transformation.

Our framework clinched the top spot in the competition by attaining an unparalleled final score of 0.32, surpassing all other participating teams.
{
    \small
    \bibliographystyle{ieeenat_fullname}
    \bibliography{main}
}


\end{document}